# Adaptive pumping for spectral control of random lasers

Nicolas Bachelard[1], Sylvain Gigan[1], Xavier Noblin[2], Patrick Sebbah[1*]

[1]Institut Langevin, ESPCI ParisTech, CNRS UMR7587, 1 rue Jussieu, 75238 Paris cedex 05, France
[2]Laboratoire de Physique de la Matière Condensée, Université de Nice-Sophia Antipolis, CNRS UMR 7336, Parc Valrose, 06108 Nice Cedex 02, France
*Correspondence to: patrick.sebbah@espci.fr

**A laser is not necessarily a sophisticated device: Pumping energy into an amplifying medium randomly filled with scatterers, a powder for instance, makes a perfect "random laser." In such a laser, the absence of mirrors greatly simplifies laser design, but control over emission directionality or frequency tunability is lost, seriously hindering prospects[1-4] for this otherwise simple laser. Lately, we proposed a novel approach to harness random lasers[5], inspired by spatial shaping methods recently employed for coherent light control in complex media[6]. Here, we experimentally implement this method in an optofluidic random laser[7] where scattering is weak and modes extend spatially and strongly overlap, making individual selection *a priori* impossible. We show that control over laser emission can indeed be regained even in this extreme case by actively shaping the spatial profile of the optical pump. This unique degree of freedom, which has never been exploited, allows selection of any desired wavelength and shaping of lasing modes, without prior knowledge of their spatial distribution. Mode selection is achieved with spectral selectivity down to 0.06nm and more than 10dB side-lobe rejection. This experimental method paves the way towards fully tunable and controlled random lasers and can be transferred to other class of lasers.**

In conventional lasers, the optical cavity determines the main characteristics of the laser emission: its geometry fixes the emission wavelengths and the lasing modes, as well as the direction of the laser emission. The active medium itself essentially provides for the optical gain when pumped by an external source. There are however situations where laser emission cannot be fully controlled by the optical cavity. This is obviously the case in cavity-less lasers such as random lasers where the optical feedback is solely provided by multiple scattering within the gain medium[1-4]. Coherent laser emission has been reported in various active random media, raising strong interest recently for this new class of lasing sources, which are easy to fabricate and have demonstrated unique properties and promising applications8. The absence of a well-defined cavity however results in an unpredictable random emission spectrum and multi-directionality. If the scattering is not strong enough to tightly confine the modes[9-12], they overlap and the situation gets even worse: strong mode competition and temporal chaotic behavior are expected[13-16], seriously limiting practical interests for random lasers.

In this letter, we achieve experimental control of the random laser spectral emission by optimization of the optical pump profile. The first proposal for laser optimization based on non-uniform pumping can be traced back to the seminal paper by Kogelnik and Shank17, where it was suggested in the conclusion that, instead of applying a spatial modulation of the index of refraction, the pump intensity could be spatially modulated to vary the gain periodically to realize a distributed feedback (DFB) laser. This idea was next applied by the same authors to tune the emission wavelength of a DFB laser18. By varying the angle between two interfering pump beams, they achieved broad tunability over a bandwidth of 64 nm. Surprisingly, besides this example, the spatial modulation of the optical pump intensity has not been used much. Most commonly, optimization of the electric pumping has been tested for instance in high-speed singlemode distributed feedback semiconductor lasers to reduce the frequency chirp responsible for the increase of the spectral linewidth of DFB lasers19. But in these systems, the structured electrodes are fixed, precluding any tunability. Very recently, selection of particular lasing modes of a micro-random laser was achieved using a specific geometry where spatial and directional selectivity of the lasing modes was still possible20. In contrast, the novel method we recently proposed5 is effective at all lasing wavelengths, in the strongly as well as the weakly scattering regime where modes extend over the whole sample and overlap spatially, preventing any mode selection by local or directional pumping as proposed in ref.[20]. We successfully tested this method numerically5 albeit with a simplified model which did not include mode competition, gain saturation and spectral fluctuations. Here, we experimentally show that active control of the pump spatial profile using a spatial light modulator is an effective method in practice that survives shot to shot fluctuations and strong nonlinear behavior. We are able to bring a random laser from multimode to singlemode operation at any selected emission wavelength even in the regime of weak scattering where mode overlap and mode competition are the strongest. Surprisingly, laser emission control via pump shaping is easily achieved despite the complexity of the problem. This opens the perspective to fully controlled random laser singlemode emission, while preventing mode competition and improving laser stability. Our method extends the paradigm shift operated recently by wavefront shaping techniques for imaging through opaque media, to the control of highly complex media, namely open disordered nonlinear active media. It shows that complexity is not necessarily an obstacle but may help to meet the optimization criterion21. Most interestingly, it turns a single random laser into a versatile laser that can shift laser characteristics (e.g. different emission wavelengths or directivity), a unique feature hard to achieve in conventional laser with a regular optical cavity.

We consider the optically-pumped one-dimensional (1D) random laser described in Fig. 1 and Methods, inspired from an optofluidic random laser we recently introduced[7] for easy integration of lasers in complex optofluidic structures[22]. The pump laser is reflected off a computer-driven spatial light modulator (SLM) to modulate its spatial intensity profile and is focused into a narrow strip line to pump the dye. The optical setup is described in more details in the Methods. Figures 2a and 2b compare the emission spectra integrated over 50 shots, I(λ), just below threshold and above threshold, respectively. Below threshold, amplified spontaneous emission result in a broad spectral peak resulting from amplified spontaneous emission (ASE). When the pump exceeds the threshold, sharp peaks emerge in the spectrum. Their measured spectral linewidth is 0.03 nm (full width at half maximum) and is not limited by the 20 pm-resolution of the imaging spectrometer used here (see Methods). The number of peaks in the frequency range 560-565 nm is 35±3 for 10 sample-realizations. This is consistent with the spectral density of lasing modes found numerically by modeling a macroscopic 1D system with geometry similar to the experiment (see Supplementary Materials). Because the random structure is static and the dye solution flows continuously within the microfluidic channel, dye-bleaching is reduced and the lasing modes are stable and reproducible over several hours. Singlemode operation is yet extremely difficult to achieve when the random laser is uniformly pumped as mode's thresholds are very close. However, we claim that the sensitivity of the emission spectrum to global and local variations of the optical pump profile[23] should enable singlemode laser operation at a selected wavelength, $\lambda_0$. To do so, we need to find the pump profile, if it exists, which maximizes the ratio $R(\lambda_0) = I(\lambda_0)/I(\lambda_1)$, where $\lambda_1$ corresponds to the lasing mode with highest intensity, apart from mode $\lambda_0$. A $R(\lambda_0)$ much larger than unity signifies that mode $\lambda_0$ has been selected and that singlemode operation at this wavelength is possible.

At this point, it is important to underline the nature of the scattering within our random laser. The small index contrast between the PDMS pillars and the dye solution, $\Delta n = 1.42 - 1.36 = 0.06$, results in weak scattering. This is confirmed numerically by computing the localization length $\xi = -(d\langle \ln T \rangle/dL)^{-1}$ from the dependence of the transmission $T$ of an ensemble of 100 random configurations with sample length $L$[23]. We find $\xi = 22$ mm at 560 nm, much longer than the sample size (2.8 mm). In this regime, the modes extend over the whole system and therefore strongly overlap with each other. Hence it is impossible to predict *a priori* the correct pump profile required to select the targeted mode, $\lambda_0$, at the expense of all others. While linear methods fail, we have shown numerically that an iterative algorithm can provide with an optimized though complex solution[5]. However, the transfer matrix model used there was only valid below threshold; it did not include gain saturation, nonlinear mode competition and laser instabilities, which are expected to be all the more significant that the random laser operates in the weakly scattering regime[10-12]. In particular shot-to-shot fluctuations and noise put experimental demonstration at a disadvantage. Here, we chose the derivative-free Simplex algorithm, which is less sensitive to experimental noise than e.g. gradient-based algorithms (used in ref. 5), to minimize the cost function, $1/R(\lambda_0)$. At each iteration, a new pump profile is computed, displayed on the SLM and applied to the pump stripe; the spectrum averaged over 50 shots is acquired and $R(\lambda_0)$ is computed. Each iteration takes about 5 seconds. After 228 iterations, the algorithm converges to an optimized pump profile. The corresponding spectrum is shown in Fig. 2c, when mode at $\lambda_0 = 561.77$ nm is targeted. We chose for demonstration a lasing mode which is not the first to lase under uniform pumping, i.e. $R(\lambda_0) = 0.4$ initially. We obtain $R(\lambda_0) = 13.1$ after optimization, which corresponds to a sideband rejection of 11.4 dB. The complete optimization cycle starting from uniform pumping and multimode emission spectrum is presented in a movie (Supplementary Materials) which shows the convergence of the algorithm to a stable pump profile and a singlemode spectrum. Figure 3a shows the convergence of $R(\lambda_0)$ together with the spatial correlation of pump profile vs. iteration number. We also check that no mode-hoping between laser lines occurs during the optimization process so that the mode selected is indeed the mode initially targeted. This is assessed *a posteriori* by following spectrally the modes as the depth of the modulation of the pump profile is increased stepwise from uniform to optimized pump profile (see Supplementary Materials). Actually, it is possible by this method to optimize almost any of the lasing modes present in the multimode spectrum measured with uniform pumping. The convergence criterion $R(\lambda_0)$ is reported in Fig. 3b before and after optimization for 15 lasing modes measured in the range 560-566 nm. Optimized $R(\lambda_0)$ ranges between 5.7 and 15.3. Optimization is efficient even away from the center of the gain curve and even for weak lasing modes with initial $R(\lambda_0)$ as low as 3%. Most remarkable, we succeeded in selecting independently two lasing modes only separated by 0.06 nm, as shown in Fig. 4. The optimization process clearly leads to two distinct pump profiles associated with two different modes at $\lambda_0 = 561.08$ nm and $\lambda_0 = 561.14$ nm. It is worth noting that mode selection is demonstrated here in the most extreme case of weak scattering: by extension, spectral control will be facilitated in any random laser with higher scattering where lasing thresholds are lower, the range of singlemode operability larger and spatial discrimination easier[5].

Gain nonlinearities such as mode competition and hole burning effect play a crucial role in multimode random lasers[15]. This is illustrated in Fig. 5 where mode intensity vs. pump fluence is shown for three different modes: the targeted mode at $\lambda_0 = 561.77$ nm, the first mode to lase under uniform pumping at $\lambda_1 = 562.50$ nm and another lasing mode at $\lambda_2 = 563.4$ nm. In Fig. 5a uniform pumping was applied and all three modes shown experience intensity saturation as a result of cross-saturation which critically limit output intensity. In contrast, the targeted mode is the first to lase when the optimized pump profile is applied as shown in Fig. 5b and its intensity increases with pump fluence well beyond the saturation level of Fig. 5a (see also Fig. 2c). Figure 5b reveals that the consequences of the pump profile optimization are three-fold: (a) lasing threshold of competing modes are significantly increased relatively to the targeted mode; (b) the slope efficiency (linear fit in Fig. 5b), which measures the conversion rate of pump energy into lasing, is considerably reduced for competing modes as compared to the targeted mode; (c) consequently, emission at $\lambda_0$ is now possible at higher output intensity than in the case of uniform pumping (Fig. 5a). These observations provide a clue to the nonlinear mechanism involved in the optimization process. We believe the pump shaping most probably redistributes optimally the gain to increase mode competition and enhance cross-saturation effects due to spatial hole burning in favor of the targeted mode. This is in contrast to a mechanism based on spatial or directional selective pumping[20] which is precluded in the regime of strong modal overlap we consider here. The mechanism we propose here calls for further theoretical investigation.

The method demonstrated here is not restricted to the control of the optofluidic random laser we have tested. It is in principle scalable down to 100 μm-sized 1D systems as we are able to preserve a well-defined intensity modulation with pixels as small as 3 μm. The extension to 2D random lasers is also possible despite the increased modal density, provided the system is small enough or index contrast high. Other emission properties such as directivity[24], brightness or pulse duration could be controlled in a similar way as we demonstrated it here for the control of the emission wavelength. What makes random laser control unique is that, thanks to its complexity, the same system can give different lasing characteristics (e.g. different output wavelengths or emission patterns) controlled by the pump. This is hard, if not impossible, to achieve with conventional lasers made with simple cavities. Beyond the control of random lasers, other classes of lasers may benefit from our method. High power semiconductor lasers for instance suffer from a loss of control at high pumping rate when strong nonlinear effects, including nonlinear refraction, mode competition and self-focusing/defocusing effects, may lead to laser instabilities and beam filamentation25. Finding the optimized spatial profile of the pump beam which improves the laser brightness, can be a precious guide to the design of patterned electrodes when switching to electric pumping. The same concept can be envisioned for electrically pumped random lasers as recently proposed in the mid-infrared regime26. Finally, a controllable random laser is a unique platform for laser physicists with many parameters to play with (geometry, openness, local or non-uniform pumping, absorption, scattering …) and to explore new questions impossible or difficult to address with other lasers (e.g. non-hermitian physics, exceptional points in lasers[27], coupled lasers, partially absorbing lasers, …)27.

**Methods**

**Optofluidic random laser preparation:** The polydimethylsiloxane (PDMS) optofluidic random lasers are fabricated following the soft lithography protocol described by Xia and Whitesides[28]. The geometric structure of the photomask shown in Fig.1 is imprinted into a 1 mm-thick negative photoresist SU-8 mold. This mold is then used to replicate the polymer microstructures. A 10:1 PDMS:cross-linker mixture (Sylgard 184) is poured into the mold and then degassed for 10 min at a few mmHg vacuum pressure, and cured at 90°C for 90 minutes. Holes are perforated in the device to create inlets and the microchannel is bonded on a glass slide by plasma treatment.

**Optofluidic random laser description:** The gain medium is a $2.5 \times 10^{-3}$ M ethanolic solution of Rhodamine 6G circulating in a 2.8 mm-long polydimethylsiloxane (PDMS) microchannel, that comprises a linear chain of randomly distributed rectangular PDMS pillars. The alternation of 70 layers with two different indices of refraction ($n_{polymer}$=1.42 and $n_{dye}$=1.36), provides multiple scattering and the necessary feedback to stimulate spontaneous emission into random laser emission. The microchannel is 60µm-wide and 15 µm-deep. The pillars are 12 µm-thick and 40 µm large. Their random positions are uniformly chosen in the range ±6µm around a periodic arrangement of 40 µm-period. Actually, the artificial spatial disorder we chose is unnecessary. Indeed, the limited accuracy of the photolithographic process results in a 1µm-tolerance in the position and thickness of the pillars, providing the necessary disorder at the optics scale, as shown in[7]. An auxiliary microchannel circulated by a separate dye flow (5.0 x $10^{-3}$M ethanolic solution of Nile Blue), serves to image laser radiation scattered out by the structured channel (Fig. 1).

**Experimental setup and procedure:** The laser beam of a doubled Q-switch Nd:YAG laser (Quantum Ultra: 532 nm, 7 ns pulse duration, maximum output energy 30mJ, repetition rate 10 Hz) is expanded to illuminate uniformly the surface of the SLM (Holoeye LC 2002). The SLM itself, which seats between crossed polarizers to work in amplitude modulation, is placed in the object plane of a telescope with x5 reduction and is imaged on the sample after compression through a $f$ = 6mm cylindrical lens. This setup provides with a 1.4 mm-long, 4 µm-large laser stripe line with nearly free-diffraction modulation down to 10 µm-large rectangular pixels. We chose to focus tightly the pump beam to a narrow line in order to enforce single transverse mode laser operation. The length of 1.4 mm has been chosen to limit the amplified spontaneous emission (ASE) and to provide a manageable modal density. We checked experimentally that doubling the length increases the ASE by a factor 3 and the number of modes by 2. The microchannel is precisely aligned with the laser stripe line under a Zeiss Axioexaminer microscope and is imaged with Hamamatsu Orca-R2 silicon CCD camera microscope. The laser emission is collected via an optical fiber connected to a Horiba iHR550 imaging spectrometer equipped with a 2400 l/mm grating and a liquid nitrogen cooled Symphony II camera (sampling rate 1MHz, 1024*56 pixels, 26 µm pixel pitch). The entrance slit is 50 µm. The resulting spectral resolution is 20 pm. The integrating time is 1 s.

**Optimization method:** The intensity modulation of the pump stripe is obtained by dividing the SLM into 32 lines coded on 256 levels, which results after recompression on the sample in 32 46µm-large pixels. Any arbitrary 1D-pump profile, $f(x)$, can be decomposed over the 32-column vectors $X_i$ of the 32x32 binary Hadamard matrix[29,30]. This basis is chosen here for two main reasons: first, it ensures the same total reflectance of the SLM for all elements (except the first one); it is therefore possible to operate near lasing threshold in the linear regime of the laser for all vectors $X_i$ of the Hadamard basis, provided we set a minimum background illumination (here of $\alpha_0 = 120$). Second, each element $X_i$ of the basis is uniformly spatially distributed on average, preserving the original multimode spectrum, while changing the spectral contribution of each mode as seen in Fig. 2a and 2b, as opposed to local pumping which would modify essentially the emission spectrum[23]. The pump profile therefore writes: $f(x) = \frac{1}{255}[\alpha_0 + \sum_{1 \to 32} \alpha_i X_i]$ where the $\alpha_i$ take discrete values in the range [0,135]. Each vector $(\alpha_i)_{i \in [1,32]}$ corresponds to a particular pump profile associated with a particular emission spectrum $I(\lambda)$. To achieve singlemode operation at a desired lasing mode, $\lambda_0$, we need to find the non-trivial $(\alpha_i)_{i \in [1,32]}$ which maximize the ratio $R(\lambda_0) = I(\lambda_0)/I(\lambda_1)$, where $\lambda_1$ corresponds to the lasing mode with highest intensity, apart from mode $\lambda_0$. Here, $I(\lambda)$ represents the intensity at $\lambda$ after spectrum baseline has been subtracted, the baseline being defined has the intensity at $\lambda$ = 555 nm.

Experimentally, 5 spectra averaged over 10 shots are acquired every second and then averaged for a given pump profile $f(x)$. Between successive acquisitions, the system is uniformly pumped ($f(x) = 1$) to clean up the gain medium from any memory effect due to possible residual thermal effect from the previous pump profile. A typical optimization lasts 30 min. After several hours, spectral shift becomes significant relative to the spectral precision of the measure, as a result of PDMS infiltration by ethanol. This explains why different data sets were collected in Fig. 3b.

We checked the robustness of the solution and found that the optimization is lost ($R(\lambda_0)$ divided by 2) when the optimized pump profile is perturbed on every pixel by a white noise with a standard deviation of 7%. The algorithm provides therefore a sharp and accurate solution, sensitive to local perturbations. We find however that the optimized solution is not unique (local optimum) but other profiles can operate single mode lasing at the same frequency, depending on the starting vertex used in the algorithm.

**Optimization algorithm:** we use the Nelder-Mead simplex (direct search) method implemented in the *fminsearch* function of Matlab. We modified this function by setting the initial step (*usual_delta* parameter in *fminsearch*) to 0.5 to explore a large region of the 32-dimension space. The number of pixels (32) has been determined for best trade-off between sensitivity and computation time. It has to be also a power of 2 for definition of the Hadamard basis.

**Acknowledgments**

We thank J. P. Huignard, S. Bhaktha, and J. Andreasen for useful discussions.

P.S., N.B. & S.G. are thankful to the LABEX WIFI (Laboratory of Excellence within the French Program "Investments for the Future") under reference ANR-10-IDEX-0001-02 PSL*.

P.S. is thankful to the ANR under Grant No. ANR-08-BLAN-0302-01 and the Groupement de Recherche 3219 MesoImage.

S.G. is funded by the European Research Council (grant number 278025).

Correspondence and requests for materials should be addressed to P. S.

**Supplementary Information**

1. Experimental tracking of the targeted lasing mode from uniform to optimized pumping.

2. 1D-numerical simulation of the experimental random laser.

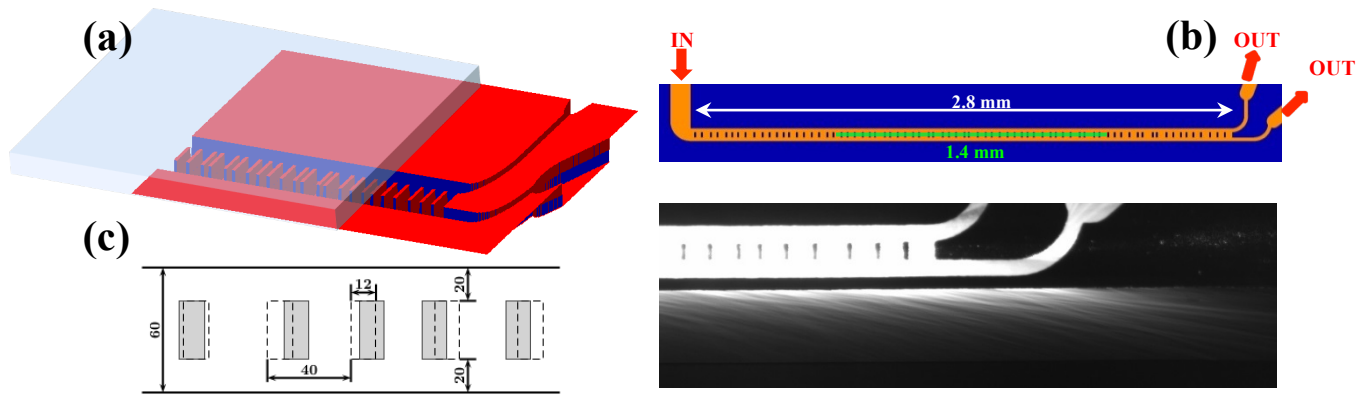

**Figure 1 : Optofluidic random laser. (a)** 3D partial view and **(b)** complete top view of the mask. **(c)** The microchannel is structured into a 1D-random distribution of 12 µm-thick rectangular pillars separated on average by 28 µm gaps. Geometric dimensions are given in *µm*. A dye solution (Rhodamine 6G in ethanol) flows through the 3mm-long structured PDMS microchannel which is plasma-bonded on a glass slide partially represented in (a). A pressure differential at the two outlets forces dye flow between the scattering pillars and prevents dye bleaching. The structure is pumped by a Q-switched Nd:YAG laser at 532 nm shaped into a 1.4 mm-long stripe line (green line in **(b)**). **(d)** Imaging of the in-plane scattered laser emission. Dye (Nile blue) solution circulates in a 80 µm-wide auxiliary microchannel parallel to and 10 µm-apart from the structured-channel and fluoresces at laser emission wavelength.

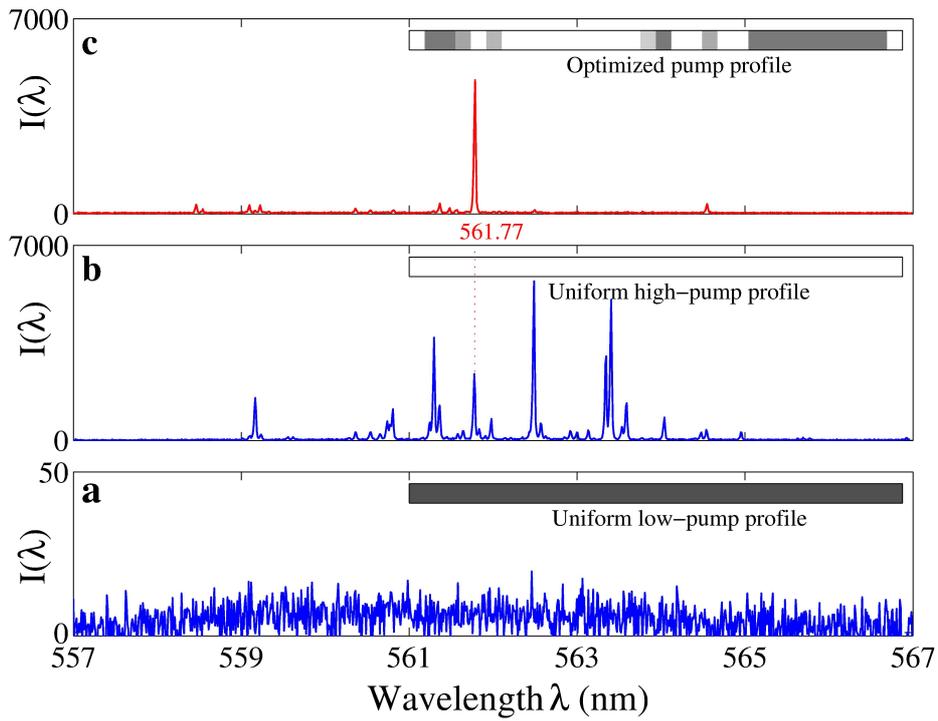

**Figure 2: From multimode to singlemode operation. (a)** Fluorescence spectrum for uniform pumping below threshold. **(b)** Laser emission spectrum for uniform pumping above threshold. **(c)** Emission spectrum after optimization process to select laser emission at $\lambda_0$ = 561.77 nm. Singlemode operation is achieved after 228 iterations. The corresponding nonuniform pump profile as displayed on the spatial light modulator (SLM) is shown in inset. The grayscale ranges from 0 (black) to 255 (white).

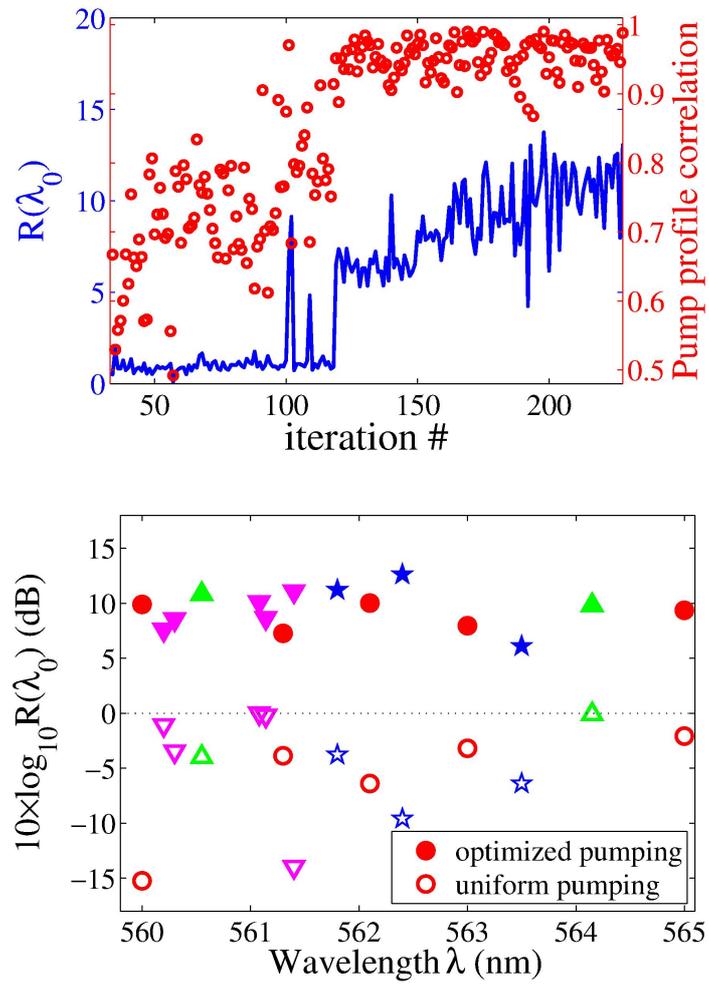

**Figure 3 : (a) Convergence of the optimization process.** Optimization of mode $\lambda_0$ = 561.77 nm. Evolution vs. iteration number of the inverse cost function $R(\lambda_0)$ (full line) and the normalized spatial correlation between iteration *i* and *i*+1 of the spatial profile of the pump (open circles). Despite large experimental fluctuations and wide exploration of the solution space, convergence is eventually reached when the modulated pump profile stabilizes. **(b) Optimization efficiency over the emission spectrum.** The optimization algorithm has been applied to select 15 lasing modes. **Open symbols**: $R(\lambda_0)<1$ before optimization. **Full symbols**: $R(\lambda_0)\gg1$ after optimization. The logarithmic representation gives $R(\lambda_0)$ in dB. Different symbols correspond to different set of experiments (see methods).

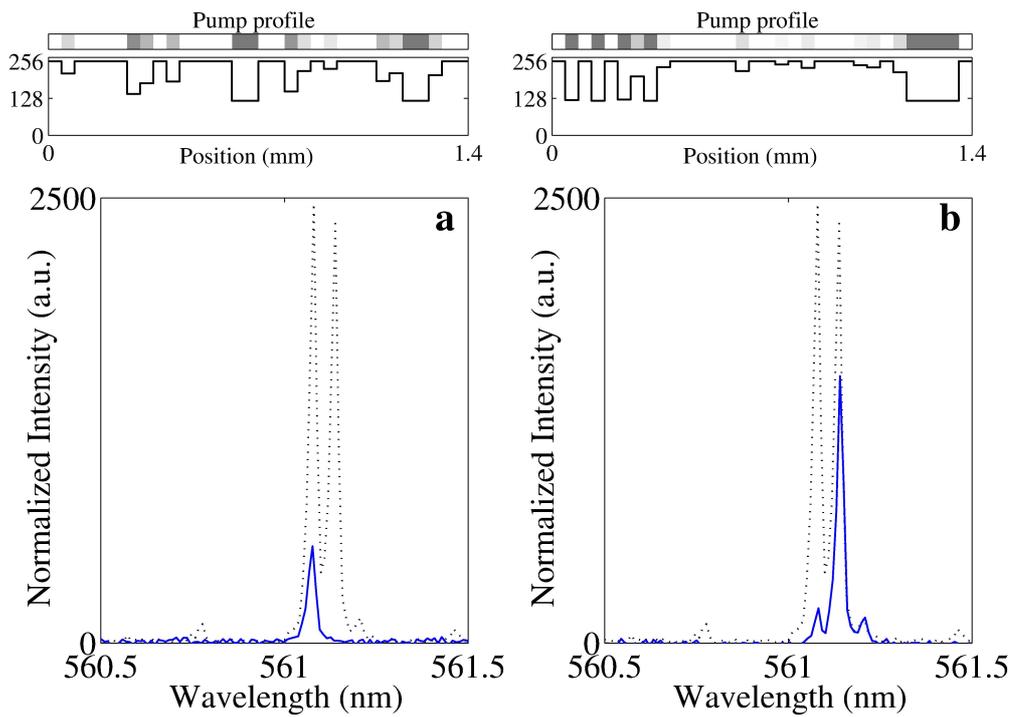

**Figure 4 : Spectral selectivity.** Optimizations at (a) $\lambda_0$= 561.08 nm and (b) $\lambda_0$= 561.14 nm. The corresponding optimized pump profile is plotted versus position and the pattern displayed on the SLM is shown in grey scale. Although the two modes are only separated by 0.06 nm, the algorithm converges to two distinct spatial pump profiles associated with two different lasing modes.

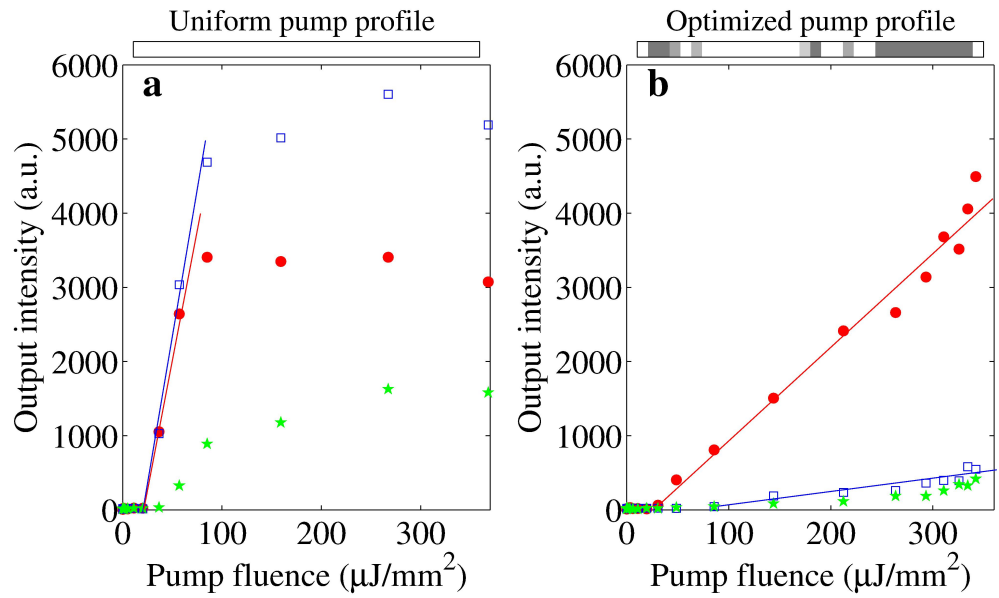

**Figure 5 : Intensity versus pump fluence.** Case of **(a)** uniform pumping and **(b)** optimized pumping for optimization routine carried out at $\lambda_0$ = 561.77 nm. The pump profile as displayed on the SLM is shown on top. **Full red circles**: the targeted mode at $\lambda_0$ = 561.77 nm; its threshold slightly increased from 21 to 28 µJ/mm². **Blue squares**: the mode which lases first under uniform pumping at $\lambda_1$ = 562.50 nm; its threshold doubled from 21 to 44 µJ/mm². **Green stars**: another mode at $\lambda_2$ = 563.4 nm. Lines are linear fit to the stimulated regimes. In (b) the x-axis shows the actual fluence impinging on the sample (which is different from the incident pump beam when pumping is not uniform).

# Supplementary Information

## 1. Experimental tracking of the targeted lasing mode from uniform to optimized pumping.

We demonstrate that the mode selected by the optimization process is the actual targeted mode of the multimode spectrum when the laser is pumped uniformly. Because, the Simplex algorithm we use in the optimization procedure as described in the main text constantly explore new regions of the parameter space, it is not possible to follow continuously the targeted mode (see fluctuations in Fig. 3a) and to assert that no "mode-hoping" occurred during the optimization process. We therefore performed the following experiment: the modulation depth of the optimized pump profile is progressively increased starting from zero (uniform pumping) until the singlemode regime is reached. Figure S1 confirms that the initially targeted mode of the uniformly-pumped spectrum can be followed continuously, and that no mode-hoping between laser lines occurs during the optimization process.

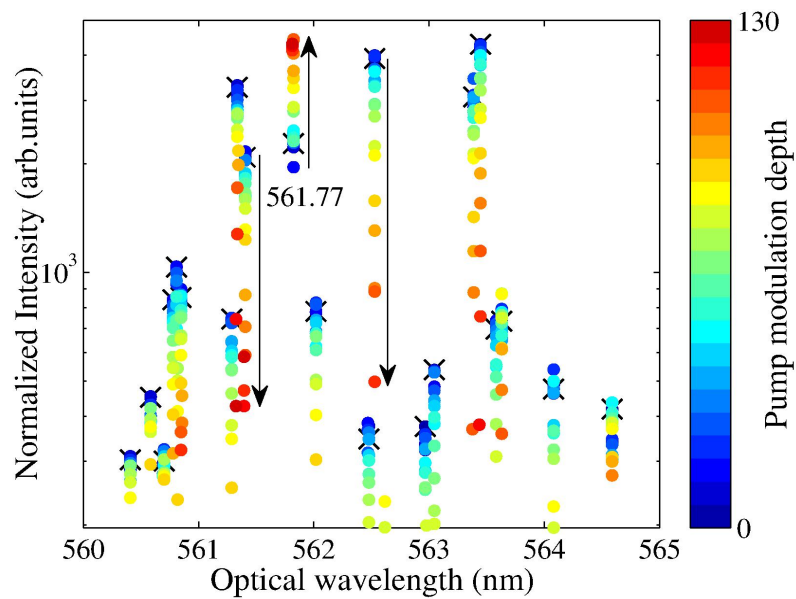

**Figure S1: Evolution in the experiment of the modes intensity with increasing pump modulation depth.** Normalized intensity vs. optical wavelength of the lasing modes measured at successive values of the pump profile modulation depth. The final pump profile modulation corresponds to the optimization of mode $\lambda_0$= 561.77 nm discussed in Fig. 2, 3 and 5. Starting from uniform pumping (dark blue), the baseline of the pump profile is progressively decreased until the optimized profile is recovered (red). The intensity of the targeted mode increases progressively without any mode-hoping while all other modes eventually stop lasing.

## 2. 1D-numerical simulation of the experimental random laser

We modeled the optofluidic random laser using physical and geometrical parameters close to the experiment. Here, we assume however the system to be strictly one-dimensional. The dye solution is modeled by a frequency-dependent gain, with a transition at $\lambda_a$= 562.5 nm and a spectral width $\Delta\lambda_a$= 30 nm (see e.g. P.R. Hammond, Optics Communications, **29**, 331 (1979)). Absorption has not been included. The transfer matrix method is used as in[5] to compute the wavelength and threshold of the lasing modes. The method is only valid below threshold: it does not include mode saturation and competition effects. The emission spectrum for uniform pumping is shown as crosses in Fig. S2. Exploring deep in the complex spectrum, we find 46 modes in the range 560-565nm for this particular configuration and 46±2 on a statistical sample. This is comparable to the modal density experiment (34±3), considering that mode competition is not taken into account in the simulation and that large threshold modes are unlikely to lase.

We checked numerically that mode selection is possible, despite the relatively high modal density (compared to the simulations in 5 where the system was smaller and hence the modal density). The lasing mode at 564.88 nm was chosen has its threshold is not the lowest. After a hundred of iterations, the targeted mode is selected with an increased threshold well below the threshold of other modes as shown in Fig. S2, which is consistent with the experimental results of Fig. 5.

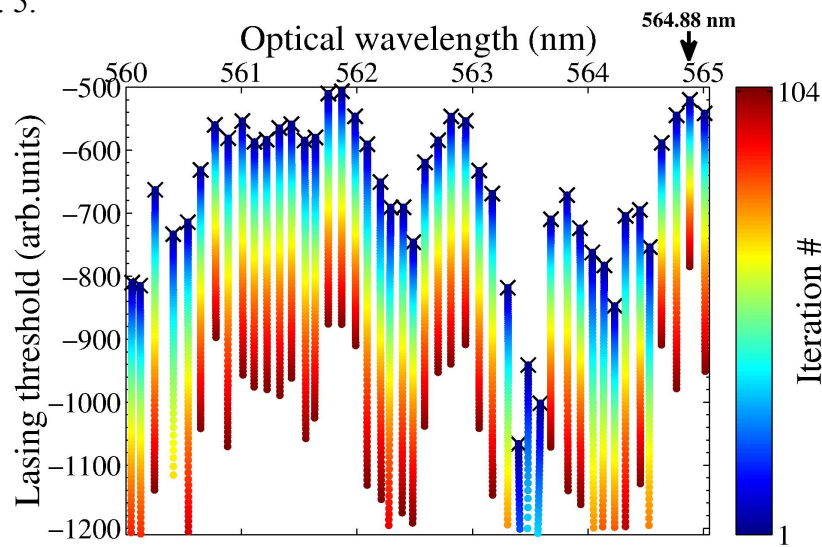

**Figure S2: Numerical simulations.** Threshold vs. optical wavelength of lasing modes in a 1D system with geometry similar to the experiment. **Crosses**: Lasing spectrum for uniform pumping. **Dots:** Lasing modes at successive iterations (from blue to red) when the optimization routine is applied to select the lasing mode at